\documentclass[aoas,MSNbibl,nameyear,seceqn,dvips]{arximspdf}
\usepackage{graphicx}

\doi{10.1214/12-AOAS551} 
\volume{6}
\issue{3}
\pubyear{2012}
\firstpage{870}
\lastpage{894}
\setattribute{copyright}{owner}{In the Public Domain}

\makeatletter
\newcommand{\bbeta}{\bolds{\beta}}
\newcommand{\bmu}{\bolds{\mu}}
\newcommand{\vx}{\mathbf{X}}
\newcommand{\bx}{\mathbf{X}}
\newcommand{\blx}{\mathbf{x}}
\newcommand{\blf}{\mathbf{f}}
\newcommand{\eqref}[1]{(\ref{#1})}
\renewcommand{\citep}[1]{\citeauthor{#1} \citeyear{#1}}
\newtheorem{thmm}{Theorem}[section]

\newtheorem{prop}{Proposition}[section]

\newproclaim{algo}{Algorithm}
\makeatother

\begin{document}
\begin{frontmatter}

\title{Functional dynamic factor models with application to yield curve forecasting}
\runtitle{Functional dynamic factor models}

\begin{aug}
\author[A]{\fnms{Spencer} \snm{Hays}\corref{}\ead[label=e1]{Spencer.Hays@pnnl.gov}},
\author[B]{\fnms{Haipeng} \snm{Shen}\thanksref{t2}\ead[label=e2]{haipeng@email.unc.edu}}
\and
\author[C]{\fnms{Jianhua Z.} \snm{Huang}\thanksref{t3}\ead[label=e3]{jianhua@stat.tamu.edu}\ead[label=u1,url]{http://www.foo.com}}
\runauthor{S. Hays, H. Shen and J. Z. Huang}
\thankstext{t2}{Supported in part NSF Grants DMS-06-06577,
CMMI-0800575 and DMS-11-06912.}
\thankstext{t3}{Supported in part by NCI (CA57030), NSF (DMS-09-07170),
and by
Award No. KUS-C1-016-04, made by King Abdullah University of Science
and Technology (KAUST).}

\affiliation{Pacific Northwest National Laboratory, University of North
Carolina at~Chapel~Hill~and Texas A\&M University}
\address[A]{S. Hays\\
Applied Statistics\\
National Security Directorate\\
Pacific Northwest National Laboratory\\
Richland, Washington 99354\\
USA\\
\printead{e1}}

\address[B]{H. Shen\\
Department of Statistics\\
\quad and Operations Research\\
University of North Carolina\\
\quad at Chapel Hill\\
Chapel Hill, North Carolina 27599\\
USA\\
\printead{e2}}

\address[C]{J. Z. Huang\\
Department of Statistics\\
Texas A\&M University\\
College Station, Texas 77843-3143\\
USA\\
\printead{e3}}
\end{aug}

\received{\smonth{3} \syear{2011}}
\revised{\smonth{9} \syear{2011}}

%
\begin{abstract}
Accurate forecasting of zero coupon bond yields for a continuum of
maturities is paramount to bond portfolio management and derivative
security pricing. Yet a universal model for yield curve forecasting has
been elusive, and prior attempts often resulted in a trade-off between
goodness of fit and consistency with economic theory. To address this,
herein we propose a novel formulation
which connects the dynamic factor model (DFM) framework with concepts
from functional data analysis: a DFM with functional factor loading
\textit{curves}. This results in a model capable of forecasting
functional time series. Further, in the yield curve context we show
that the model retains economic interpretation. Model estimation is
achieved through an expectation-maximization algorithm, where the time
series parameters and factor loading curves are simultaneously
estimated in a single step. Efficient computing is implemented and a
data-driven smoothing parameter is nicely incorporated. We show that
our model performs very well on forecasting actual yield data compared
with existing approaches, especially in regard to profit-based
assessment for an innovative trading exercise. We further illustrate
the viability of our model to applications outside of yield forecasting.
\end{abstract}

\begin{keyword}
\kwd{Functional data analysis}
\kwd{expectation maximization algorithm}
\kwd{natural cubic splines}
\kwd{cross-validation}
\kwd{roughness penalty}.
\end{keyword}
%
\end{frontmatter}

\section{Introduction}\label{secintro}
The yield curve is an instrument for portfolio management and for
pricing synthetic or derivative securities [\citet{dl06}]. Bond prices
are hypothesized to be a function of an underlying continuum of yields
as a function of maturity, known as the yield curve. Our contribution
to the yield literature is pragmatic: we introduce a dynamic factor
model with functional coefficients which reconciles the theory-based
desire to model yield data as a curve with the applied need of
accurately forecasting that curve over time.

The yield curve is a \textit{theoretical} construct not without its own
inherent \textit{practical} difficulties. First and foremost, although
yield determines prices, only bond \textit{prices} are observed for a set
of discrete maturity horizons; from these a corresponding discrete set
of yields are calculated. Thus, the yields themselves are not directly
observed, nor is an entire curve for every possible maturity. Further,
not only is it of interest to know the yield for all maturities at each
point in time (cross-sectional), but also for a single maturity as it
evolves over time (dynamic). Finally, because a bond at time $i$ of
maturity~$t$ is essentially the same bond as the one at time $i+1$ of
maturity $t-1$, there is also a certain amount of systematic \textit
{cross-correlation} in yield data. Therefore, predictive modeling of
bond data needs to consider each of the cross-sectional, dynamic and
cross-correlational behaviors.

To this end, yield curve models have traditionally assumed either of
two formulations. The first is theoretical in nature: as in \citet
{Hull1990} and \citet{Heath1992}, for a given date the emphasis is on
fitting a yield curve to existing yields based on no-arbitrage
principles stemming from economic theory. The other approach is the
so-called equilibrium or affine-class models where time series
techniques are used to model the dynamics of yield on a short-term or
instantaneous maturity, and yields for longer maturities are then
derived using an affine model. This method has been developed in works
such as \citet{Vasicek1977}, \citet{Cox1985}, and \citet{Duffie1996}.

These contrasting methods illustrate the dichotomy of yield forecast
models. As a practical matter, goodness of fit is paramount in a model
for it to be of any use. Still, a yield model should be consistent with
its underlying theory, and maintain a degree of economic
interpretation. Cross-sectional/no-arbitrage models ignore the dynamics
of yields over time [as noted in \citet{dl06}, \citet{koop10}, e.g.] and thus
threaten the former yet satisfy the latter. Time series/equilibrium
models place emphasis on the former at the expense of the latter [as
seen in \citet{duffee02}].

What we propose in this paper is a synthesis of the cross-sectional and
dynamic considerations mentioned above. We approach yield curves as a
\textit{functional} time series; the yields of the observed maturities
are a discrete sampling from a true underlying yield \textit{curve}. To
this end, we conflate concepts from functional data analysis [FDA;
\citeauthor{rams02} (\citeyear{rams02,rams05})] and from dynamic factor analysis/modeling [DFM;
\citet{basil94}, e.g.]. Ours is a dynamic factor model with \textit{functional
coefficients} which we call (not surprisingly) the functional dynamic
factor model (FDFM). These functional coefficients, or \textit{factor
loading curves}, are natural cubic splines (NCS): a significant result
which facilitates interpolation of yields both within and out of sample
so that forecasts are indeed true yield curves. While the factor
loadings account for the cross-sectional/curve dimension of yields, the
dynamic factors, in turn, determine the evolution of these functions
over time. Thus, they account for the time series and
cross-correlational nature of yield data. Our particular specification
of the FDFM enables its estimation via the Expectation Maximization
(EM) algorithm [\citet{demp77}].

Why the need for \textit{both} a functional and a dynamic factor
framework? Recall that the unifying goal is to develop a model that is
consistent with the concept of the yield \textit{curve} posited by
economic theory and is of use for practical forecasting. A naive
attempt to merge the latter need with the former is to model yields for
all observed maturities over time as a~multivariate time series.
However, as the number of observed maturities increases to even
moderate size, vector autoregressive models (VARs)---for example---become
intractable because of high dimensionality.

Abstracting from the yield setting for a moment, in a more general
sense large multivariate time series have been successfully modeled
[\citet{engle81}, \citet{gewe81},
\citet{mol85}, \citet{pena87}, \citet{pena04}, to name just a few]
using a dynamic factor approach. In DFMs the multivariate data are
assumed to be dependent on a small set of unobserved dynamic factors.
This solves the dimensionality problem, yet DFMs \textit{per se} leave to
question the interpretability of the unobserved factors. Further, in
our present context, DFMs fall short of producing a functional yield
\textit{curve}.

To incorporate the functional aspect, we propose to combine the DFM
framework with ideas from functional data analysis (FDA). However, FDA
in general is an area still nascent in development, and most
applications deal primarily with collections of \textit{independent}
curves. Earlier work by \citet{besse00} applied functional
autoregressive models (FAR) to univariate climatological data: the
seasonal cycle is hypothesized to be functional. In a similar
hypothesis, \citet{shen09} forecasted periodic call volume data using a
method akin to functional principle component analysis (FPCA). In an
applied setting more similar to ours, \citet{Hyndman09} developed a
weighted FPCA method to forecast time series of curves and applied it
to multivariate time series of fertility or mortality data indexed by
different ages. Yet, unlike these models where FPCA and time series
modeling are performed in separate steps, ours is a method that
estimates both functional and time series components \textit
{simultaneously}, and does so in a quite natural manner.

Within the context of yield curve forecasting, other recent
developments have begun to reconcile the statistical viability of DFMs
and functional data analysis with the underlying theory in regard to
yield dynamics---a~constraint which all but requires the usually
absent interpretation for the dynamic factors. \citet{dl06} introduced
the Dynamic Nelson--Siegel model (DNS): a three factor DFM with
functional coefficients estimated in two steps, which extends the
original Nelson--Siegel model [\citet{NS1987}]. The functional
coefficients are pre-specified as fixed parametric curves and the
authors further provide an economic interpretation of each. \citet
{koop10} extended the DNS specification to allow (G)ARCH volatility
and a fourth dynamic factor which allows time dependence to the
otherwise fixed parametric factor loading curves. Another DFM-type
approach is provided by \citet{bows08} which present a cointegrated DFM
using natural cubic splines (NCS). Spline knots serve as dynamic
factors following an error correction model process; the knot locations
are determined via an
initial exhaustive search-selection procedure prior to model
estimation. As noted in \citet{koop10}, cointegrated factors present a
difficulty in terms of retaining economic interpretation.

Presented in this paper is our functional dynamic factor model (FDFM)
which we show to perform very well in regard to yield curve
forecasting. Further, we do so in multiple assessments which highlight
the model's capability of accurately forecasting the entire function as
well as the potential profit generated from employing these forecasts
in trading strategies. Finally, via our online supplement (a~brief
description follows Section~\ref{secconc}), in simulation studies we
illustrate the accuracy of both FDFM forecasts and predicted
parameters. In either sense the FDFM outperforms existing models which
require either multiple-step estimation or lack a functional component.

It is worth noting our FDFM is in a similar vein as those of the
aforementioned yield models: a dynamic factor model with functional
coefficients; one which---quite coincidentally---even exploits the
properties of NCS for the cross-sectional/curve dimension of yields.
However, unlike \citet{dl06}, the FDFM functional coefficients are
estimated; thus, they are free to vary with the particular application
to explain the functional nature of the data. Further, as opposed to
the existing two classes of models, estimation of the FDFM is achieved
in a single step. Within the yield context it will be seen that the
FDFM satisfies our two aforementioned criteria: goodness of fit and
economic interpretability. That the factor loading curves are estimated
facilitates application of the FDFM to contexts outside of yield curve
forecasting as well. We will show through simulation (online
supplement) that our specification even permits the inclusion of
observed nonlatent variables in the dynamic factors similar to \citet{DRA06}.

The remainder of our paper is organized as follows. In Section \ref
{secfdfm} we develop our model, including discussion of its
formulation, details regarding estimation and significant results in
terms of application and utilization. Section~\ref{secappl} examines
in detail the motivating example of real yield data in multiple
forecasting and assessment exercises. Finally, we conclude with Section
~\ref{secconc} containing a discussion of our key findings and some
directions of future research. In an online supplement we illustrate
simulation results and highlight the model's viability for both
forecasting and parameter accuracy, especially in regard to
applications outside of yield curve forecasting. In addition, our
online supplement [\citet{hays2011online}] provides technical proofs for
the theorem and propositions presented in Section~\ref{secfdfm}.

\section{Functional dynamic factor models} \label{secfdfm}
Abstracting for a moment from the present setting of yield curve
forecasting, consider the more general process of a time series of
curves $\{x_i(t)\dvtx t\in\mathcal{T};i=1,\ldots,n\}$, where $\mathcal{T}$
is some continuous interval and $i$ indexes discrete time. It is
hypothesized that each curve is composed of a forecastable smooth
underlying curve, $y_i(t)$, plus an error component, $\varepsilon_i(t)$,
that is,
%
\begin{equation}\label{eqy+e}
x_i(t) = y_i(t) + \varepsilon_i(t).
\end{equation}
There are two primary goals of a functional time series model: to
provide an accurate description of the dynamics of the series, and to
accurately forecast the smooth curve $y_{n+h}(t)$ for some forecast
horizon $h > 0$.

In practice, of course, only a discrete sampling of each curve is
observed. Specifically, consider a sample of discrete points $\{
t_{1},t_{2},\ldots,t_{m}\}$ with $t_{j}\in\mathcal{T}$ for $j \in\{
1,\ldots,m\}$.
The observed data for the $i$th curve are $x_{ij} \equiv x_i(t_j)$,
$j\in\{1,\ldots,m\}$.

\subsection{The model} \label{ssecmodel}
By synthesizing DFM and FDA, we propose a model referred to as the
functional dynamic factor model (FDFM). The formulation is similar to
that of a DFM where the observed data $\{x_{ij}\}$ is a function of a
small set of $K$ latent dynamic factors $\{\beta_{ik}; k=1,\ldots,K\}$
and their corresponding factor loadings. But in this setting the factor
loadings $f_{kj} \equiv f_k(t_j)$ are discrete samples from continuous,
unobserved though nonrandom factor loading curves $f_{k}(\cdot)$.
Together, the dynamic factors with their functional coefficients
generate the forecastable part of the time series of curves $\{
x_{i}(t)\}$.

In theory, the dynamic factors can follow any type of time series
process such as (V)ARIMA, but for the purpose of this paper we focus on factors
which are independent, stationary $\operatorname{AR}(p)$ processes. Although it is not
necessary for the number of lags $p$ to be the same for each factor, as
a matter of notational convenience we simply define $p=\max{\{
p_1,\ldots
,p_K\}}$ and use the appropriate placement of zeros.
The factors can include explanatory variables\setcounter{footnote}{2}\footnote{These could be
economic indicators or seasonal effects, for example.} or just a constant.
In the former case, we have a $1 \times d$ regressor vector $A_{ik}$
having the $d \times1$ coefficient vector~$\mu_{k}$. Similarly, we let
$d=\max{\{d_1,\ldots,d_K\}}$,
but do retain the option for the regressors themselves to differ among
factors; thus, we continue to use the $k$ subscript per factor.
Finally, for the model to be identified, we require that the functional
coefficients are orthonormal.\footnote{Other types of constraints may be
employed to ensure identification, such as conditions on the covariance
function of the factor loading curves.} The model is explicitly stated as
%
\begin{equation} \label{eqfdfm}
\cases{
\displaystyle x_{i}(t_{j}) = \sum_{k=1}^{K}\beta_{ik}f_{k}(t_{j}) + \varepsilon
_{i}(t_{j}), \vspace*{2pt}\cr
\displaystyle \beta_{ik}-A_{ik}\mu_{k}=\sum_{r=1}^{p} \varphi_{rk} (\beta
_{i-r,k}-A_{i-r,k}\mu_{k}) + v_{ik},\vspace*{4pt}\cr
\displaystyle \int_{T}f_{k}(t)f_{l}(t)\,dt = \cases{
1, &\quad $\mbox{if } k = l,$\vspace*{2pt}\cr
0, & \quad $\mbox{otherwise,}$}}
\end{equation}
with $\varepsilon_i(t_j) \equiv\varepsilon_{ij}\stackrel{\mathrm{i.i.d.}}{\sim}
N(0,\sigma^2)$, $v_{ik}\stackrel{\mathrm{i.i.d.}}{\sim} N(0,\sigma_{k}^2)$ and
$E[v_{ik}\varepsilon_{i'j}]=0$ for $i,i'=1, \ldots, n$. Should we require
only a constant in place of regressors, then $A_{ik}\mu_{k}$ is a~scalar $\mu_{k}$ for all $i$.
With the assumption of stationarity, this
yields the constant $c_k=\mu_k (1-\sum_{r=1}^{p}\varphi_{rk})$. This is
a broad framework that includes the standard versions of both DFMs and
FPCA models: when the coefficients $\{f_k(t)\}$ are nonfunctional,
model \eqref{eqfdfm} reduces to the standard DFM; when the factors
$\{
\bbeta_k\}$ are nondynamic, the model is similar to FPCA.

\subsection{Estimation} \label{ssecestimation}
With the error assumptions for model \eqref{eqfdfm}, we propose
estimation via maximum likelihood (ML). To ensure smooth and functional
estimates for the factor loading curves, we augment the likelihood
expression with ``roughness'' penalties [\citet{green94}] and maximize
a~\textit{penalized} log-likelihood expression.
Because our dynamic factors are unobserved, we consider this a problem
of missing data, and use the expectation maximization (EM) algorithm
[\citet{demp77}] to estimate model parameters and smooth curves.

\subsubsection{Penalized likelihood} \label{sssecPL}
Let the $n\times m$ matrix $\bx$ denote collectively the observed data
where the $(i,j)$th element of $\bx$ is $x_{ij}$ for $i=1,\ldots,n$,
$j=1,\ldots,m$. Each row of $\bx$ corresponds to a yield curve for a
fixed date; each column represents the time series of yield for a
specific maturity. Next, we denote $ f_{kj}=f_k(t_j)$, the $m \times1$
vector $\blf_k' = [f_{k1},\ldots,f_{km}]$, and the factor loading curve
matrix \textbf{F} as
\[
\mathbf{F}' = [\mathbf{f}_{1}, \ldots, \mathbf
{f}_{K}],
\]
so that the \textit{rows} of \textbf{F} are the transposed column vectors
$\blf_k$ [this convention is to conform with some standard factor
analysis matrix notation; see \citet{basil94}, e.g.].
In a similar manner, we define $\bbeta_{k} = [{\beta_{1k}\enskip \cdots\enskip\beta
_{nk}}]'$ and the matrix $\mathbf{B}_{n \times K} = [{\bbeta_{1}\enskip
\cdots\enskip
\bbeta_{K}}]$. Thus, the columns of $\mathbf{B}$ are the time series
factors $\bbeta_1, \ldots, \bbeta_K$. Then, the model \eqref{eqfdfm}
is represented in matrix form as
%
\begin{equation} \label{eqmatrixmodel}
\mathbf{X}_{n \times m} = \mathbf{B}_{n \times K} \mathbf{F}_{K
\times
m} + \bolds{\varepsilon}_{n \times m} = \sum_{k=1}^K \bbeta_k
\blf_k'
+ \bolds{\varepsilon},
\end{equation}
where $\bolds{\varepsilon} = [\varepsilon_{ij}]_{n \times m}$ with
$\varepsilon_{ij}=\varepsilon_i(t_j)$.

Assuming the matrix of dynamic factors $\mathbf{B}$ is observable, the
log-likelihood expression can be obtained by successive conditioning of
the joint distribution for~\textbf{X} and \textbf{B}:
%
\begin{equation} \label{eqjointdist}
l(\mathbf{X},\mathbf{B}) = l(\mathbf{B}) + l(\mathbf{X}|\mathbf{B}).
\end{equation}
Because we have assumed that the $K$ factors of $\operatorname{AR}(p$) series are
independent, their joint distribution is the product of the individual
distributions. To each of those, we further condition on the first $p$
values of each factor time series; thus, our likelihood \eqref{eqjointdist} is a \textit{conditional} one. For ease of notation we assume
there are no regressors in the factor time series. Then
%
\begin{equation} \label{eqlkhdBs}
 l(\mathbf{B}) = (n-p)\!\sum_{k=1}^{K}\ln(2\pi\sigma_{k}^{2})
+\!\sum_{i=p+1}^{n}\sum_{k=1}^{K}\frac{1}{\sigma_{k}^{2}}
\Biggl(\beta_{ik}-c_{k}-\!\sum_{r=1}^{p}\varphi_{rk}\beta_{i-r,k}\Biggr)^{2},\hspace*{-35pt}
\end{equation}
and
%
\begin{equation} \label{eqlkhdxs}
l(\mathbf{X}|\mathbf{B}) = nm\ln(2\pi\sigma^{2}) + \frac
{1}{\sigma
^{2}}\sum_{i=1}^{n}\sum_{j=1}^{m}\Biggl(x_{ij}-
\sum_{k=1}^{K}\beta_{ik}f_{kj}\Biggr)^{2}.
\end{equation}

To ensure the underlying factor loading curve $f_k(\cdot)$ is smooth,
following \citet{green94},
we introduce roughness penalties to \eqref{eqlkhdxs} to obtain the
following penalized log-likelihood:
%
\begin{eqnarray} \label{eqPL}
l_p(\mathbf{X},\mathbf{B}) &=& l(\mathbf{B}) + l_p(\mathbf
{X}|\mathbf
{B}),
\nonumber
\\[-8pt]
\\[-8pt]
\nonumber
\nonumber&\equiv& l(\mathbf{B}) + \Biggl[l(\mathbf{X}|\mathbf{B}) +
\sum_{k=1}^{K}\lambda_{k}\int[f_{k}''(t)]^2 \,dt\Biggr].
\end{eqnarray}
The penalty parameter $\lambda_k$ controls how strictly the roughness
penalty is enforced, and we allow it to differ for each loading curve
(thus the ``$k$'' subscript).
The selection process for the penalty parameters is discussed in
Section~\ref{sseccom}. We refer to the latter term in equation
 \eqref
{eqPL}, $l_p(\bx|\mathbf{B})$, as the penalized sum of squares (PSS).
Intuitively, optimization of PSS balances a familiar goodness-of-fit
criterion with a smoothness requirement for the resulting estimates of $f_k(t)$.

Below we assume the dynamic factors are known and discuss how to
estimate the AR model parameters and the smooth factor loading
curves.\vadjust{\goodbreak}

When the dynamic factors have no regressors the conditional MLEs for
the AR parameters ($\{\sigma_k^2,c_k,\varphi_{1,k},\ldots,\varphi
_{p,k}\}$) are the same as the ordinary least squares (OLS) solutions.
In the case where the factors do have regressors, an additional step is
required to alternatively solve for the AR parameters $\{\varphi
_{1,k},\ldots,\varphi_{p,k}\}$ and the regressor coefficient vectors
$\{
\mu_k\}$. The resulting solutions are the (feasible) generalized least
squares (GLS) solution; see \citet{judge85} for a detailed discussion.
We do consider this general formulation in the simulation studies
reported in our online supplement; a brief discussion follows Section
~\ref{secconc}.

Now we discuss how to estimate the loading curves $f_k(t)$. In order to
allow the curves to have their own smoothness, through allowing
different $\lambda_k$, we proceed in a sequential manner to estimate
$f_k(t)$ one at a time, incorporating penalty parameter selection for
that loading curve through cross-validation, as discussed in Section
~\ref{sseccom}.

According to Theorem 2.1 of \citet{green94}, for fixed $k$, the
minimizer $\hat{f}_k(\cdot)$ of PSS is a natural cubic spline with knot
locations $t_1,\ldots,t_m$. Further, this NCS interpolates the discrete
vector $\hat{\blf}_k$ which is the solution to the minimization problem
%
\begin{equation} \label{eqminprobA}
\min_{\blf_k} [l(\mathbf{X}|\mathbf{B}) + \lambda_{k} \blf_k'
\bolds{\Omega} \blf_k ],
\end{equation}
where $\bolds{\Omega}_{m \times m}$ is a matrix determined solely
by the spline knot locations; the explicit formulation of $\bolds{\Omega}$ is deferred until Section~\ref{ssecNCS}.

Let $\mathbf{X} \equiv \operatorname{vec}(\mathbf{X})$ which stacks the columns
of $\mathbf{X}$ into an $nm \times1$ vector. Then using the Kronecker
product $\otimes$, model \eqref{eqmatrixmodel} can be rewritten in
vector form as
%
\begin{equation}\label{eqvecmodel}
\cases{\displaystyle
\vx= (\mathbf{F}' \otimes\mathbf{I}_{n})\bbeta+ \operatorname{vec}(\bolds{\varepsilon})
=\sum_{k=1}^K (\blf_k \otimes\mathbf{I}_n)\bbeta_k +
\operatorname{vec}(\bolds{\varepsilon}), \vspace*{2pt}\cr
\displaystyle\vx= \sum_{k=1}^K (\bbeta_k \otimes\mathbf{I}_m)\blf_k +
\operatorname{vec}(\bolds{\varepsilon}).}
\end{equation}
The lattermost form facilitates a straightforward derivation of the
optimal factor loading curves. To see this,
consider the solution $\hat{\blf}_k$ for fixed \mbox{$k \in\{1,\ldots,K\}
\equiv\mathbb{K}$}. For the remaining $h \in\mathbb{K}$, we define
$\vx
^{*} = \vx- \sum_{h \neq k} (\bbeta_h \otimes\mathbf{I}_m)\blf_h$.
Then the minimization problem \eqref{eqminprobA} is equivalent to
%
\begin{equation}\label{eqminprobB}
\min_{\mathbf{f}_{k}} \biggl\|\frac{1}{\sigma}\vx^{*} - \frac
{1}{\sigma
}(\bbeta_k \otimes\mathbf{I}_m) \cdot\blf_k\biggr\|^2 + \lambda
_k\mathbf{f}_{k}'\bolds{\Omega}\mathbf{f}_{k},
\end{equation}
where $\|\cdot\|$ is the Euclidean norm. Expanding the first term and
differentiating with respect to $\blf_{k}$ yields the solution
%
\begin{equation} \label{eqridgesoln}
\hat{\blf}_{k} = \frac{1}{\sigma^2}\biggl[\frac{\|\bbeta_k\|
^2}{\sigma
^2}\mathbf{I}_{m} + \lambda_k\bolds{\Omega} \biggr]^{-1}
(\mathbf{I}_{m} \otimes\bbeta_{k}' )\vx^{*},\vadjust{\goodbreak}
\end{equation}
or $\sigma^{-2} \mathbf{S}(\mathbf{I}_{m} \otimes\bbeta_{k}' )\vx^{*}$
for $\mathbf{S} \equiv[\frac{\|\bbeta_k\|^2}{\sigma
^2}\mathbf
{I}_{m} + \lambda_k\bolds{\Omega} ]^{-1}$; $\mathbf{S}
\equiv\mathbf{S}(\lambda_k)$.
In Section~\ref{sseccom} we derive a generalized cross-validation
(GCV) procedure for the selection of each~$\lambda_k$.

\subsubsection{EM algorithm} \label{sssecEM}
In the realistic situation that $\mathbf{B}$ is unobservable, we treat
it as missing data and resort to the EM algorithm for maximizing the
observed data log-likelihood.
First, the EM is inaugurated with initial values for the factors and
factor loading curves. From these initial values, maximum likelihood
estimates for the remaining parameters from $\Theta$ are calculated
based on equations \eqref{eqlkhdBs}, \eqref{eqlkhdxs} and~\eqref{eqPL};
we call this \textit{Step} 0. Then the algorithm
alternates between the E-step and the M-step. In the \textit{E-step},
values for the factor time series are calculated as conditional
expectations given the observed data and current values for the MLEs.
In the \textit{M-step}, MLEs are calculated for the factor loading curves
and other parameters based on the factor scores from the conditional
expectations in the E-step. After the initial step, the E-step and the
M-step are repeated until differences in the estimates from one
iteration to the next are sufficiently small. More details are given below.

\textit{Step} 0:
Akin to the method used in \citet{shen09}, initial values for \textbf
{B} are composed of the first $K$ singular values and left singular
vectors from the singular value decomposition (SVD) of the data matrix
\textbf{X}. Initial values for \textbf{F} are the corresponding right
singular vectors. From these, initial parameter estimates are computed
for~$\sigma^2$ and the set of factor parameters $\{\sigma
_k^2,c_k,\varphi_{1,k},\ldots,\varphi_{p,k}\}$.

\textit{The E-step}: Derivation of the conditional moments for the
E-step requires the expressions of some of the unconditional moments.
Define the $n \times n$ variance matrix for $\bbeta_k$ as $\bolds{\Sigma}_{k}$,
and let \textbf{c} be the $K \times1$ vector with
elements $c_k/[1-(\sum_{r=1}^{p}\varphi_{r,k})]$. Then, using equations
 \eqref{eqvecmodel},
%
\begin{eqnarray}
E[\bbeta] &\equiv&\bmu_{\bbeta} = \mathbf{c} \otimes\mathbf{1}_n, \qquad
E[\vx] \equiv\bmu_{\mathbf{X}} = (\mathbf{F}' \otimes\mathbf
{I}_{n})\bmu_{\bbeta},\nonumber\\
\operatorname{Var}[\bbeta]& \equiv&\bolds{\Sigma}_{\bbeta} = \operatorname{diag}\{\bolds{\Sigma}_{1},\ldots,\bolds{\Sigma}_{K}\},
\nonumber
\\[-8pt]
\\[-8pt]
\nonumber
 \operatorname{Cov}[\bbeta,\vx]
&\equiv&
\bolds{\Sigma}_{\bbeta,\mathbf{X}} = \bolds{\Sigma
}_{\bbeta
}(\mathbf{F} \otimes\mathbf{I}_{n}),
\\
\operatorname{Var}[\vx] &\equiv&\bolds{\Sigma}_{\mathbf{X}} =
(\mathbf
{F}' \otimes\mathbf{I}_{n})\bolds{\Sigma}_{\bbeta}(\mathbf{F}
\otimes\mathbf{I}_{n})+\sigma^{2}\mathbf{I}_{nm}.\nonumber
\end{eqnarray}

Next, using properties of multivariate normal random vectors, the
conditional distribution of $\bbeta|\bx$ can be found. Let
%
\[
\pmatrix{
\bbeta\vspace*{2pt}\cr
\vx}
\sim
N\left[
\pmatrix{
\bmu_{\bbeta}\vspace*{2pt}\cr
\bmu_{\mathbf{X}}
}
,
\pmatrix{
\bolds{\Sigma}_{\bbeta} & \bolds{\Sigma}_{\bbeta
,\mathbf
{X}}\vspace*{2pt}\cr
\bolds{\Sigma}_{\mathbf{X},\bbeta} & \bolds{\Sigma
}_{\mathbf{X}}}
\right].
\]
Then
%
\begin{eqnarray} \label{eqconddist}
\cases{
\displaystyle\bmu_{\bbeta|\mathbf{X}}\equiv E[\bbeta|\vx] = \bmu_{\bbeta}+
\bolds{\Sigma}_{\bbeta,\mathbf{X}}\bolds{\Sigma
}_{\mathbf{X}}^{-1}
(\vx-\bmu_{\mathbf{X}}), \vspace*{2pt}\cr
\displaystyle\bolds{\Sigma}_{\bbeta|\mathbf{X}}\equiv
\operatorname{Var}[\bbeta|\vx] = \bolds{\Sigma}_{\bbeta}-
\bolds{\Sigma}_{\bbeta,\mathbf{X}}\bolds{\Sigma
}_{\mathbf
{X}}^{-1}\bolds{\Sigma}_{\mathbf{X},\bbeta},\vspace*{2pt}\cr
\displaystyle E[\bbeta\bbeta'|\bx] = \bolds{\Sigma}_{\bbeta
|\mathbf{X}}
+\bmu_{\bbeta|\mathbf{X}} \bmu_{\bbeta|\mathbf{X}}'.}
\end{eqnarray}

From a computational standpoint there is concern over the inversion of
$\bolds{\Sigma}_{\mathbf{X}}$ which is of order $nm$. Because the
EM is an iterative procedure, this could be especially problematic.
However, we can use the following result based on the
Sherman--Morrison--Woodbury factorization [\citet{numrec92}, e.g.] to
simplify the computation:
\begin{prop} \label{thSx}
%
\begin{eqnarray} \label{eqsxfacinv}
\bolds{\Sigma}_{\mathbf{X}}^{-1}&=& \sigma^{-2}\mathbf
{I}_{nm} -
\sigma^{-4}(\mathbf{F}'\otimes\mathbf{I}_{n})
[\sigma^{-2}\mathbf{I}_{nK}+ \bolds{\Sigma}_{\bbeta
}^{-1}]^{-1}
(\mathbf{F}\otimes\mathbf{I}_{n}).
\end{eqnarray}
\end{prop}

A derivation of this result is included in our online supplement; the
form of the result is not so important as what it means. Instead of
inverting $\bolds{\Sigma}_{\mathbf{X}}$ directly, which is an $nm
\times nm$ matrix, only the middle matrix\vspace*{1pt} $[\sigma^{-2}\mathbf
{I}_{nK}+ \bolds{\Sigma}_{\bbeta}^{-1}]$ needs to be
inverted. This matrix is of smaller size $nK \times nK$. Further, as
$\bolds{\Sigma}_{\bbeta}$ is block diagonal, then $\sigma
^{-2}\mathbf{I}_{nK}+ \bolds{\Sigma}_{\bbeta}^{-1}$ is as well.
Thus, using this factorization, the inversion of $\bolds{\Sigma
}_{\mathbf{X}}$ is reduced from an $nm \times nm$ inversion to~$K$, $n
\times n$ inversions.

With the conditional moments, the E-step of the EM posits that the
missing data (the time series factors) are replaced with the known
values of the conditional distribution given \textbf{X}. Thus,
in the following M-step, in solving for the MLEs, expressions involving
$\bbeta_k$ will utilize values from $\bmu_{\bbeta|\mathbf{X}}$,
$\bolds{\Sigma}_{\bbeta|\mathbf{X}}$ and $E[\bbeta\bbeta
'|\bx]$.

\textit{The M-step}: For each EM iteration, the M-step optimizes the
conditional penalized log-likelihood in equation \eqref{eqPL} given
the observed data and the current parameter estimates for $\Theta$. It
is clear from equations \eqref{eqlkhdBs} and~\eqref{eqlkhdxs}
that in the MLEs, the factor time series appear either singly or in
terms of cross products both within and between factors. Values for
terms like~$\beta_{ik}$ come directly from the vector $\bmu_{\bbeta
|\mathbf{X}}$. But because a term like~$\beta_{ik'}\beta_{hk}$,
$k,k'=1,\ldots,K$, $i,h = 1,\ldots,n$, is a conditional expectation of
a product, its replacement values are obtained from the $E[\bbeta
\bbeta
'|\vx]$ matrix. We will show in Section~\ref{sseccom} some rather
fortunate results to simplify computation of the conditional
expectation of the factor products.

The M-step, then, is just a matter of making these substitutions into
the likelihood, and solving for the MLEs. After the M-step, we return
to the E-step to update the values for the factor time series. This
procedure is repeated until the parameter estimates from one iteration
of the EM are sufficiently close to those of the next.

\subsection{Connection with natural cubic splines} \label{ssecNCS}
We now explain the origin of the penalty matrix $\bolds{\Omega}$
from equation \eqref{eqminprobA} following \citet{green94}. Let
$h_j = t_{j+1} - t_j$. For $j=1,\ldots,m$, we define the banded matrix
$Q_{m \times(m-2)}$ with columns numbered in a nonstandard way:
elements $q_{jj'}$ denote the $j=1,\ldots,m$th row and $j'=2,\ldots
,(m-1)$st column of $Q$.\vadjust{\goodbreak} These elements, in particular, for $|j-j'| < 2$,
are given by
%
\begin{equation} \label{eqQelem}
q_{j-1,j} = h^{-1}_{j-1}, \qquad q_{jj} = -h^{-1}_{j-1} - h^{-1}_{j}, \qquad
q_{j+1,j} = h^{-1}_{j},
\end{equation}
and are 0 otherwise. Further, we define the symmetric matrix $R_{(m-2)
\times(m-2)}$ with elements $r_{jj'}; j,j'=2,\ldots,(m-1)$ such that
$r_{jj'} = 0$ for $|j-j'| \geq2$ and otherwise
%
\begin{eqnarray} \label{eqRelem}
\cases{
\displaystyle r_{jj} = \tfrac{1}{3}(h_{j-1} - h_j), &\quad $\mbox{for } j=2,\ldots,m-1,$
\vspace*{2pt}\cr
\displaystyle r_{j,j+1} = r_{j+1,j} = \tfrac{1}{6}(h_{j-1} - h_j), & \quad$\mbox{for }
j=2,\ldots,m-2.$}
\end{eqnarray}

Note that $R$ is diagonal dominant and, thus, it is positive definite
and invertible. Let
%
\begin{equation} \label{eqOmdefn}
\bolds{\Omega} = QR^{-1}Q'.
\end{equation}
The following result is based on Theorem 2.1 of \citet{green94}.

\begin{prop}\label{thNCS}
For fixed k, the $\hat{f}_k(\cdot)$ optimizing \textit{PSS} in \eqref
{eqPL}
is a~natural cubic spline with knot locations at $t_j$, and
\[
\int[f_{k}''(t)]^2 \,dt =
\mathbf{f}'_k \bolds{\Omega} \mathbf{f}_k.
\]
\end{prop}

A proof of Proposition~\ref{thNCS} is included in our online Supplement.

\subsection{Forecasting and curve synthesis} \label{ssecfore}
Recall that the goal of our Functional Dynamic Factor Model (FDFM) is
to provide forecasts of an \textit{entire} curve from an observed time
series of sampled curves. Once the FDFM has been estimated, it is a
straightforward exercise to do just this. Further, due to the
functional nature of the model, we are not restricted to forecasts for
only the observed knot locations; the natural cubic spline (NCS)
results of Section~\ref{ssecNCS} allow us to forecast to any degree
of fineness between knot locations. Indeed, Proposition~\ref{thNCS}
even allows within sample imputation of an entire time series.

Forecasting is straightforward: for illustrative purposes, suppose we
estimate our FDFM with $K$ factors following an $\operatorname{AR}(1)$ process with
constants $\{c_k\}$, $k=1,\ldots,K$. Then the $h$-step ahead forecasted
curve\vspace*{-1pt} $\hat{x}_{n+h|n}(t)$ is based on the components of the forecast
of the factor time series $\hat{\beta}_{n+h|n,k}$ and the estimated
factor loading curves $\hat{f}_{k}(t)$:
%
\begin{equation}\label{eqfore}
\cases{
\displaystyle\hat{x}_{n+h|n}(t) = \sum_{k=1}^{K}\hat{\beta}_{n+h|n,k}\hat
{f}_{k}(t),\vspace*{2pt}\cr
\displaystyle\hat{\beta}_{n+h|n,k}= \hat{c}_k + \hat{\varphi}_{k} \hat{\beta
}_{n+h-1,k}
= \sum_{r=0}^{h-1}\hat{\varphi}^r\hat{c}_k + \hat{\varphi}_{k}^h
\beta_{nk}.
}
\end{equation}

The NCS result of Section~\ref{ssecNCS} ensures that $\hat{f}_{k}(t)$
is indeed a function rather than a discrete set of points. Thus, we can
interpolate $\hat{f}_{k}(t)$ to any degree of fineness between any two
knot locations $t_j$ and $t_{j+1}$.

Specifically, consider $t\in[t_{j},t_{j+1}]; j= 1,\ldots,m$. We can
compute values for an entire time series $\{\hat{x}_{1}(t)\}_{i=1}^{n}$
because each $\hat{f}_k(t)$ is an NCS. Denote $\gamma_{kj} \equiv
f_k''(t_j)$. It can be shown [\citet{green94}]
%
\begin{eqnarray} \label{eqinterF}
\hat{f}_k(t)&=& \frac{(t-t_j)f_{k,j+1} + (t_{j+1}-t)f_{kj}}{h_j}
\nonumber\hspace*{-35pt}
\\[-4pt]
\\[-12pt]
\nonumber
&&{} +
\frac{1}{6}(t-t_j)(t_{j+1}-t)\biggl[\biggl(1+\frac{t-t_j}{h_j}\biggr)\gamma_{k,j+1}
+\biggl(1+\frac{t_{j+1}-t}{h_j}\biggr)\gamma_{kj}\biggr]\hspace*{-35pt}
\end{eqnarray}
for each $k=1,\ldots,K$. For $t < t_1$, or $t > t_m$, the $\hat
{f}_k(t)$ is a linear extrapolation, which may or may not perform well
depending on whether the linearity assumption beyond the boundary knots
is suitable for the application of interest; we illustrate this
limitation in Section~\ref{sssecccurveass}. Using this method
together with equations \eqref{eqfore}, we can just as easily impute
\textit{and} forecast at the same time, a result that enables, for
example, yield forecasts for bonds of maturities that \textit{have not
been observed}.

\subsection{Computational efficiency} \label{sseccom}
This section presents results intended to ease some of the
computational aspects of the estimation for the functional dynamic
factor model, the reason for this being that the EM algorithm is an
iterative procedure and each iteration is rife with large matrix
inversions and manipulations. Further, given the results of Section
~\ref{sssecPL}, we propose to sequentially solve for each factor
loading curve $\blf_k$; $k=1,\ldots,K$.
Finally, the smoothing parameter $\lambda_k$ needs to be selected in a
data-adaptive manner for each $k$. Below, we present a (generalized)
cross-validation (GCV) procedure to achieve this. Efficient
implementation allows us to easily evaluate the GCV score over many
candidate values of $\lambda_k$.

\textit{GCV selection}:
In general, cross-validation is based on sequentially leaving out
sections of the observed data, estimating a model for each
``leave-out'' and computing some metric for how well the model predicts
the left out sections.
Although a popular method for GCV in FDA is row/curve deletion, because
the present setting involves a dynamic system of curves, deletion of a
curve removes an entire time point from the data and destroys the time
dependency structure. Therefore, here, we pursue a GCV criterion based
on a leave-out of each \textit{series} or column. In either sense, it is
costly to re-estimate the model when each of $m$ columns or~$n$ rows of
the data $\mathbf{X}$ are deleted,
for each candidate value of $\lambda_k$ and for each $k$. Fortunately
we have the following result that obviates
re-estimation of the FDFM for each column leave-out:\looseness=1
\begin{thmm} \label{thcv,gcvthm}
Let $\mathbf{X}^{*} \equiv\mathbf{X} - \sum_{h \neq k} \bbeta_k
\mathbf
{f}_k'$. Then the GCV criterion for each~$\lambda_k$ based on column
deletion is explicitly expressed by components of estimation on the
complete data:
%
\begin{equation} \label{eqSXbGCV}
\operatorname{GCV}(\lambda_k) = \frac{\|(\mathbf{I}_{m}-{\|\bbeta
_k\|^2}/{\sigma^2}\mathbf{S})(\mathbf{X}^{*})'\bbeta_k \|
^{2}/m}{[1-\operatorname
{tr}({\|\bbeta_k\|^2}/{\sigma^2}\mathbf{S})/m]^{2}}.
\end{equation}
\end{thmm}

The proof of Theorem~\ref{thcv,gcvthm} is found in our online supplement.
GCV($\lambda_k$) is calculated over a grid of possible values during
the M-step of each EM iteration for each factor loading curve. The
smoothing parameter that corresponds to the least value of GCV($\cdot$)
is selected as the optimal one. It is worthwhile to note that this can
be a computationally intensive procedure: calculating GCV($\lambda$)
for several values for $\lambda$ during each EM iteration and for each factor.
Criterion \eqref{eqSXbGCV}\vspace*{-1pt} depends on the inversion of the matrix
$\mathbf{S}^{-1}=[\frac{\|\bbeta_{k} \|^{2}}{\sigma
^{2}}\mathbf
{I}_{m} + \lambda_{k}\bolds{\Omega} ]$.
Using the\vspace*{1pt} eigen-decomposition of $\bolds{\Omega}$, a method exists
for which the only inversion required is the inversion of a diagonal
matrix. Consider the following proposition, the derivation of which is
included in our online supplement:
\begin{prop}\label{thomegadecomp}
Given the eigen-decomposition of the $m \times m$ penalty matrix
$\bolds{\Omega} = \bolds{\Gamma\Delta\Gamma}'$ with
$\bolds{\Delta}_{m\times m} = \operatorname{diag}\{\delta_{j}\}_{j=1}^{m}$, then
\[
\mathbf{S}(\lambda_{k}) =\bolds{\Gamma}\cdot
\operatorname{diag}\biggl\{\biggl(\frac{\|\bbeta_{k}\|^{2}}{\sigma^{2}}+\lambda
_{k}\delta_{j}\biggr)^{-1}\biggr\}
\bolds{\Gamma}',
\]
and
\[
 \operatorname{ tr}\{\mathbf{S}(\lambda_{k})\}=\sum_{j=1}^{m}
\frac{1}{{\|\bbeta_{k}\|^{2}}/{\sigma^{2}}+\lambda_{k}\delta_{j}}.
\]
\end{prop}

Thus, a single eigen-decomposition, followed by a diagonal matrix
inversion for each of the factors, circumvents performing an $m\times
m$ inversion for each of the~$K$ factors and each of the candidate
values for $\lambda_{k}$.

\textit{Block diagonality}: In the M-step, when products of the factors
appear, such as $\langle\bbeta_k,\bbeta_h\rangle= E[\langle\bbeta
_k,\bbeta_h\rangle|\vx]$, then the imputation comes from the
$E[\bbeta
\bbeta'|\bx]$ matrix. It can be shown that $\bolds{\Sigma
}_{\bbeta|\bx}$ is block diagonal; this property facilitates a rather
convenient result regarding between-factor cross products (the
derivation of this result is found in our online supplement).
\begin{prop}\label{prop4}
$\bolds{\Sigma}_{\bbeta|\bx}$ is block diagonal with $K$
$n \times n$ blocks. Further, for $h \neq k$, $E[\langle\bbeta
_{k},\bbeta_{h}\rangle|\vx] = \langle\bmu_{\bbeta_{k}|\bx
},\bmu
_{\bbeta_{h}|\bx}\rangle$.
\end{prop}

Therefore, the conditional expectation of a product of two (distinct)
factors is simply the product of their individual expectations. This
greatly simplifies the \mbox{M-step} calculations.

\section{Application to yield curve data}\label{secappl}

\subsection{Yield curve data}\label{ssecdata}
In this section we consider the application of our functional dynamic
factor model to actual yield data. We use the same data set as \citet
{dl06} which consists of a sample of monthly yields on zero coupon
bonds of eighteen different maturities (in months):
\[
1.5,3,6,9,12,15,18,21,24,30,36,48,60,72,84,96,108,120,
\]
from the period January 1985 through December 2000 (192 months),
originally obtained from forward rates provided by the Center for
Research in Securities Prices (CRSP), then converted to unsmoothed
Fama--Bliss yield rates [see \citet{bliss87} for the conversion methodology].

%
\begin{figure}

\includegraphics{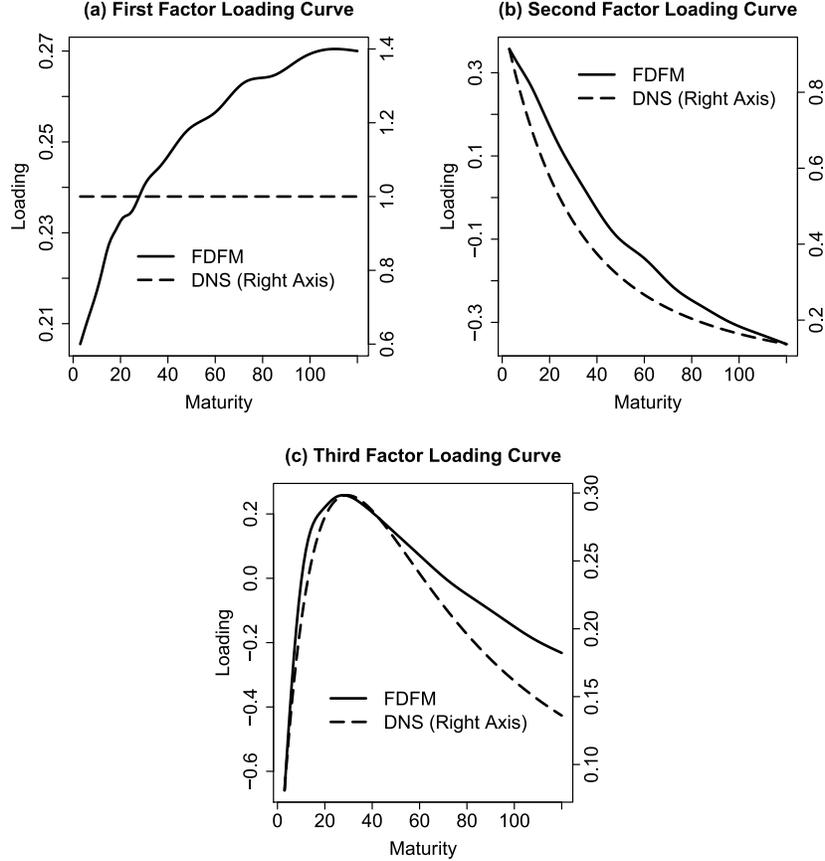}

\caption{Example of factor loading curves: FDFM curves (solid, left
axis) estimated from the period May 1985 to April 1994; pre-specified
DNS curves (dashed, right axis). FDFM estimates closely resemble the
shape of the DNS curves for the second and third factors, while the
first FDFM factor loading curve resembles a typical yield curve shape.
Dual axes are used to account for difference in scale: FDFM is
represented on the left axis; DNS on the right axis.} \label{figdsnperf}
\end{figure}

\subsection{The dynamic Nelson--Siegel model (DNS)}\label{ssecothmod}
In the following sections we compare the FDFM with the DNS model
presented in \citet{dl06}. Their model is composed of three factors
with corresponding factor loading curves. The factor loadings are
pre-specified parametric curves (see the dashed curves in Figure~\ref
{figdsnperf}) based on financial economic theory. Let $x_{i}(t)$
denote the yield at date $i$ on a zero coupon bond of maturity $t$,
then the DNS model is represented as
%
\begin{eqnarray}\label{eqDNSFs}
\cases{
\displaystyle x_{i}(t) = \sum_{k=1}^{3}\beta_{i,k}f_{k}(t) +\varepsilon_i(t) \qquad \mbox{for }
i=1,\ldots,n,\vspace*{2pt}\cr
\displaystyle f_{1}(t) \equiv1 ,\qquad
f_{2}(t) \equiv\frac{1-\exp(-\alpha_{i} t)}{\alpha_{i} t}
,\vspace*{2pt}\cr
f_{3}(t) \equiv f_2(2) - \exp(-\alpha_{i} t), &\vspace*{2pt}\cr
\displaystyle \beta_{i,k} = c_{k} + \varphi_{k}\beta_{i-1,k} +
\zeta_{i,k} \qquad
\mbox{for } k=1,2,3,}
\end{eqnarray}
evaluated at maturities $t_j$, $j=1,\ldots,m$. The first loading curve
$f_1(t)$ is constant and intended to represent the long-term component
of yields (level); the second $f_2(t)$ represents a short-term
component, or slope. Finally, the third loading $f_3(t)$ represents a
mid-term component, or curvature. The parameter $\alpha_i$ determines
the point $t^*(\alpha_i)$ at which $f_3(t)$ achieves its
maximum. While this can be estimated as a fourth factor [see, e.g.,
\citet{koop10}], \citet{dl06} set $\alpha_i$ to a fixed value for all
$i=1,\ldots,n$. This results in entirely predetermined, parametric
curves. The specific value $\alpha= 0.0609$ is determined by their
definition of ``mid-term'' as $t=30$ months.

Estimation of the DNS model is a two-step procedure. First, time
series of factor scores of $\hat{\beta}_{i,k}$ are estimated by
ordinary least squares (OLS) of $x_{i}(t_j)$ on $[1,
f_{2}(t_j),f_{3}(t_j)]$ for $j=1,\ldots,m$ at each time point
$i=1,\ldots,n$. Second, an $\operatorname{AR}(1)$ model is fit on each series $\hat
{\beta
}_{i,k}$ for the purpose of forecasting $\hat{\beta}_{n+1,k}$ and
ultimately $\hat{x}_{n+1}(t_j)$ via equation \eqref{eqfore} from
Section~\ref{ssecfore}.

\subsection{Assessment} \label{ssecforeass}
We assess the performance of the FDFM in three distinct exercises. The
first two are traditional error based assessments of forecasts or
within-sample predictions of yield curves or sections thereof. The
final application is a combination of both forecasting and curve
synthesis. Through an adaptation of the trading algorithms introduced
in \citet{bows08}, we develop trading strategies based on the forecasts
of the FDFM and DNS models and assess the resulting profit generated by each.

For each of these, as a comparison, we use the DNS specification
aforementioned above in Section~\ref{ssecothmod}. For the purpose of
making an unbiased comparison, we use a similar formulation of our FDFM
model with 3 factors following independent $\operatorname{AR}(1)$ processes. The key
distinction between this FDFM model and the DNS model is that the FDFM
estimates the model simultaneously: the smooth factor loading curves
\textit{and} the $\operatorname{AR}(1)$ parameters are estimated in a single step. In
contrast, the estimation for the DNS model requires two steps given the
\textit{pre-specified} factor loading curves: first the factor time
series are estimated; from these the $\operatorname{AR}(1)$ parameters are
determined.\looseness=-1

The key distinction between the two models raises an interesting
question: How do the factor loading curves between the two models
compare? Figure~\ref{figdsnperf}, panels~(a)--(c), show an example
of the factor loading curves estimated by the FDFM (solid line) for
the period May 1985 through April 1994. Pictured alongside, the dashed
line plots the DNS model curves. Recall the DNS motivation for the
form of $f_1$, $f_2$ and $f_3$ was an economic argument, while the
formulation of the FDFM described in Section~\ref{secfdfm} is based
entirely on statistical considerations.
Despite this, we see that the
FDFM model is flexible enough to adapt to a specific
application. Factor loading curves $f_2(t)$ and $f_3(t)$ from the FDFM
assume the behavior of those from the DNS model without imposing any
constraints that would force this. Thus, the FDFM inherits the
economic interpretation of $f_2(t)$ and $f_3(t)$ set forth in
 \citet{dl06}.
In the case of $f_1(t)$, the FDFM version resembles a
typical yield curve shape as opposed to a constant value for DNS;
however, inspecting the magnitude suggests that departure of the FDFM
version from a constant value is small.
Less typical yield curve shapes are usually characterized by deviations
in the short and mid-term yields from the norm.
This is exactly what $f_2(t)$, $f_3(t)$ and their corresponding scores capture.
Thus, we consider the first factor as the mean yield, while the second
and third account for short and mid-term deviations from this norm.

\subsubsection{Forecast error assessment} \label{sssecmsfeass}
In this section we compare the FDFM and DNS models using a rolling
window of 108 months to forecast the yield curve 1, 6 or 12 months
ahead. For example, for the one month ahead forecast we fit the models
on the first 108 months of data and forecast the 109th month, then fit
the models on the 2nd through 109th month and forecast the 110th month,
etc. Yields on bonds of maturity less than three months are omitted in
order to match the methodology used in \citet{dl06}. To compare the
models, we use the mean forecast error (MFE), root mean squared
forecast error (RMSFE) and mean absolute percentage error (MAPE):
\begin{eqnarray*}
\operatorname{MFE}_j&=&\sum_{i=1}^{r}\frac{[x_{n+h}(t_j)-\hat
{x}_{n+h}(t_j)]}{r},\\
\operatorname{RMSFE}_j& =& \sqrt{\sum_{i=1}^{r}\frac{[x_{n+h}(t_j)-\hat
{x}_{n+h}(t_j)]^{2}}{r}},
\\
\operatorname{MAPE}_j& =& \frac{100}{r}\sum_{i=1}^{r}\frac{
|x_{n+h}(t_j)-\hat{x}_{n+h}(t_j)|}{x_{n+h}(t_j)},
\end{eqnarray*}
where $r=84,79,73$ is the number of rolling forecasts for forecast
horizon $h=1,6,12$, respectively.

A summary of the forecasting performance is shown in Table~\ref{tbdsnbias}.
For month ahead forecasts, the MFE is lower (in magnitude) with
the FDFM for four out of the five displayed maturities (highlighted in
\textit{bold}), while RMSFE is lower for all five. For six months
ahead, DNS outperforms FDFM just 2 out of five times in both MFE and
RMSFE. For twelve month ahead forecasts, DNS outperforms FDFM in MFE
for 3 of 5 displayed maturities. However, FDFM has lower RMSE for all 5
maturities. In terms of MAPE, the FDFM exhibits lower MAPE than DNS
nearly uniformly for the displayed maturities and for 1, 6 and 12 month
ahead forecasts.

\begin{table}
\caption{MFE, RMSFE and MAPE: 1, 6 and 12 month ahead yield curve
forecast results. The better result between the two models is
highlighted in \textit{bold}. For 1 month ahead forecasts, the FDFM
results in lower (magnitude) MFE for most maturities, but results are
mixed for 6 and 12 months ahead. RMSFE and MAPE is typically lower with
the FDFM for 1, 6 and 12 months ahead}\label{tbdsnbias}
\begin{tabular*}{\textwidth}{@{\extracolsep{\fill}}lcccccc@{}}
\hline
&\multicolumn{2}{c}{\textbf{1 month ahead}}&\multicolumn{2}{c}{\textbf{6 months}}&\multicolumn{2}{c@{}}{\textbf{12
months}}\\[-6pt]
&\multicolumn{2}{c}{\hrulefill}&\multicolumn{2}{c}{\hrulefill}&\multicolumn{2}{c@{}}{\hrulefill}\\
\textbf{Maturity} & \textbf{DNS} & \textbf{FDFM} & \textbf{DNS} & \textbf{FDFM} & \textbf{DNS} & \textbf{FDFM}\\
\hline
&\multicolumn{6}{c@{}}{MFE}\\
3 months & $-$0.045 & \phantom{0.}\textbf{0.026}& \phantom{0.}\textbf{0.123} & \phantom{0.}0.172 & \phantom{0.}\textbf{0.203} & \phantom{0.}0.257 \\
1 year & \phantom{0.}\textbf{0.023} & \phantom{0.}0.035 & \phantom{0.}0.177 & \phantom{0.}\textbf{0.168} & \phantom{0.}0.229 &\phantom{0.}\textbf{0.215} \\
3 years & $\bolds{-}$0.056 & \phantom{0.}\textbf{0.015}& \phantom{0.}\textbf{0.022} & \phantom{0.}0.060 & \phantom{0.}\textbf{0.003} & \phantom{0.}0.013 \\
5 years & $-$0.091 & \textbf{$\bolds{-}$0.004}& $-$0.079 & \textbf{$\bolds{-}$0.021} & $-$0.166 &\textbf{$\bolds{-}$0.133} \\
10 years & $-$0.062 & \textbf{$\bolds{-}$0.023}& $\bolds{-}$0.139 & \textbf{$\bolds{-}$0.121} &\textbf{$\bolds{-}$0.316} & $-$0.318 \\[3pt]
&\multicolumn{6}{c}{RMSFE}\\
3 months & \phantom{0.}0.176 & \phantom{0.}\textbf{0.164}& \phantom{0.}\textbf{0.526} & \phantom{0.}0.535 & \phantom{0.}0.897 &\phantom{0.}\textbf{0.867} \\
1 year & \phantom{0.}0.236 & \phantom{0.}\textbf{0.233}& \phantom{0.}\textbf{0.703} & \phantom{0.}0.727 & \phantom{0.}0.998 &\phantom{0.}\textbf{0.967} \\
3 years & \phantom{0.}0.279 & \phantom{0.}\textbf{0.274}& \phantom{0.}0.784 & \phantom{0.}\textbf{0.775 }& \phantom{0.}1.041 &\phantom{0.}\textbf{0.947} \\
5 years & \phantom{0.}0.292 & \phantom{0.}\textbf{0.277}& \phantom{0.}0.799 & \phantom{0.}\textbf{0.772 }& \phantom{0.}1.078 &\phantom{0.}\textbf{0.953} \\
10 years & \phantom{0.}0.260 & \phantom{0.}\textbf{0.250}& \phantom{0.}0.714 & \phantom{0.}\textbf{0.697} & \phantom{0.}1.018 &\phantom{0.}\textbf{0.921} \\[3pt]
&\multicolumn{6}{c}{MAPE}\\
3 months & \phantom{.}2.58& \phantom{.}\textbf{2.50} &\phantom{.}8.21 &\phantom{.}\textbf{8.11} &12.99\phantom{.} &\textbf{12.05}\phantom{.} \\
1 year & \phantom{.}3.37 &\phantom{.}\textbf{3.30} &\textbf{10.25}\phantom{.} &10.29\phantom{.} &12.70\phantom{.} &\textbf{12.08}\phantom{.} \\
3 years & \phantom{.}3.79 &\phantom{.}\textbf{3.77}& 11.68\phantom{.}& \textbf{11.33}\phantom{.}& 14.16\phantom{.}& \textbf{12.71}\phantom{.} \\
5 years & \phantom{.}3.88 &\phantom{.}\textbf{3.81}& 11.94\phantom{.}& \textbf{11.42}\phantom{.}& 14.93\phantom{.}& \textbf{13.41}\phantom{.} \\
10 years & \phantom{.}\textbf{3.24}& \phantom{.}\textbf{3.24}& 10.49\phantom{.}& \phantom{.}\textbf{9.99}& 14.22\phantom{.}&\textbf{13.22}\phantom{.} \\
\hline
\end{tabular*}
\end{table}

\subsubsection{Curve synthesis} \label{sssecccurveass}
Because each factor loading curve $\hat{f}_k(\cdot)$ is an NCS, between
any two observed maturities $t_j$ and $t_{j+1}$, we can calculate the
value for $\hat{f}_k(t)$. It follows, then, that between any two time
series of yields $\{x_{i}(t_j)\}_{i=1}^{n}$ and $\{x_{i}(t_{j+1})\}
_{i=1}^{n}$, we\vspace*{1pt} are able to replicate an entire time series for the
intermediate maturity $t$: $\{\hat{x}_{i}(t)\}_{i=1}^{n}$.

\begin{figure}

\includegraphics{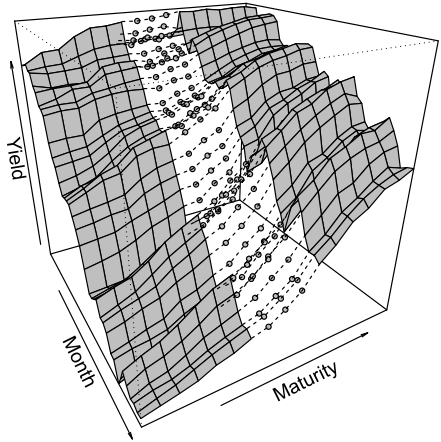}

\caption{Example of curve synthesis: Entire time series of yields are
omitted from estimation, then ``filled in'' using the imputation
described in Section \protect\ref{ssecfore}. Here, 3 consecutive maturities
have been omitted, resulting in 3 missing time series corresponding to
these maturities.} \label{figccurveex}
\end{figure}

To illustrate this point, we use the entire data set (see introduction
of Section~\ref{secappl}), that is, use $i=1,\ldots,n=192$ months of
yield data for maturities $t_j$, $m=18$. For both the DNS and FDFM
models, we delete a set of adjacent time series from the data, estimate
the model, then assess the prediction error of the predicted series in
reference to the actual deleted series.
Specifically, for our data matrix $\bx_{n \times m}$ with columns
$\blx
_1,\ldots,\blx_m$, we omit $l=1,\ldots,L$ consecutive columns from
$\bx$,
then estimate the model on the remaining $Q\equiv m-L$
maturities. From this we compute the~$L$ time series of missing data:
$\hat{\blx}_j,\ldots,\hat{\blx}_{j+L}$;
an example for the case where $L=3$ is shown in Figure~\ref{figccurveex}.
For each choice of $L$, we delete a ``horizontally'' rolling window of
width $L$ maturities and estimate the model on the remaining $Q$
maturities, $R\equiv m-L+1$ times. As an example, for $L=3$, we can
estimate the models on $\blx_4,\ldots,\blx_m$ and predict $\hat
{\blx}_1,\ldots,\hat{\blx}_3$; then estimate the models on $\blx_1,\blx
_5,\ldots,\blx_m$ and predict $\hat{\blx}_2,\ldots,\hat{\blx}
_4$, etc.

We examine the RMSFE for the $l$th omitted maturity of the $r$th
sequence; $r=1,\ldots,R$; $l=1,\ldots,L$. Because the models are
estimated based on a rolling window of maturities, for each choice of
$L$ a time series $\blx_j$ for yield $t_j$ will be estimated multiple
times. Therefore, for each choice of $L$ we take the mean of the RMSFE
of the predicted series for each maturity. We further average over our
definitions of short ($t \in[1.5,21)$), mid ($t \in[21,36]$) and
long-term ($t \in(36,120]$) horizons. Finally, we average over all
maturities as a one-number summary. These results are presented in
Table~\ref{tbrmspe} with FDFM as a fraction of DNS. Because
prediction for the FDFM model outside the range of the data is linear
extrapolation,\footnote{This is due to the NCS framework; see Section
~\ref{ssecNCS} for details.} we expect these to become increasingly
inaccurate as $L$ grows large. Thus, results are also presented
excluding extrapolated predictions in order to better illustrate the
truly functional predictions of the FDFM.

In general, as $L$ increases from 1 to 8 we see the expected decline in
the performance of the FDFM model relative to DNS. In panel (a) of
Table~\ref{tbrmspe} the average RMSFE on short-term bonds for the
FDFM remains surprisingly robust as we delete more and more maturities.
On mid-term bonds, DNS results in lower prediction error when the
number of deleted series reaches~5 or more. For long term, DNS more or
less outperforms FDFM across the board (this trend will be echoed in
Section~\ref{sssecportass}). These results are similar whether or
not the extrapolated results are included. Perhaps the best summary is
the last column in each of panel (a) and~(b) of Table~\ref{tbrmspe},
where, beyond 3 or 4 omitted maturities, the parametric based DNS model
begins to outperform the FDFM.

\begin{table}
\caption{Average RMSFE; FDFM as a fraction of DNS: \textup{(a)} with
extrapolation \textup{(b)} without extrapolation}\label{tbrmspe}
\begin{tabular*}{\textwidth}{@{\extracolsep{\fill}}lcccccccc@{}}
\hline
&\multicolumn{4}{c}{\textbf{(a) With extrapolation}} & \multicolumn{4}{c@{}}{\textbf{(b) Without
extrapolation}}\\[-6pt]
&\multicolumn{4}{c}{\hrulefill} & \multicolumn{4}{c@{}}{\hrulefill}\\
\textbf{Omitted} & \textbf{Short} & \textbf{Mid} & \textbf{Long} & \textbf{All} & \textbf{Short} & \textbf{Mid} & \textbf{Long} & \textbf{All} \\
\hline
1 & \textbf{0.88} & \textbf{0.97} & 1.05 & \textbf{0.95} & \textbf
{0.84} & \textbf{0.97} & 1.04 & \textbf{0.94} \\
2 & \textbf{0.95} & \textbf{0.90} & 1.13 & \textbf{1.00} & \textbf
{0.90} & \textbf{0.90} & 1.01 & \textbf{0.94} \\
3 & \textbf{0.94} & \textbf{0.98} & 1.06 & \textbf{0.98} & \textbf
{1.00} & \textbf{0.98} & \textbf{1.00} & \textbf{0.99} \\
4 & \textbf{0.87} & \textbf{0.93} & 1.64 & 1.14 & \textbf{0.99} &
\textbf{0.93} & 1.07 & 1.01 \\[3pt]
5 & \textbf{0.99} & 1.01 & \textbf{0.88} & \textbf{0.95} & 1.07 & 1.01
& \textbf{0.99} & 1.03 \\
6 & \textbf{0.99} & \textbf{1.00} & 1.67 & 1.26 & 1.05 & \textbf{1.00}
& 1.20 & 1.09 \\
7 & 1.34 & 1.11 & \textbf{0.93} & 1.13 & 1.24 & 1.11 & 1.16 & 1.19 \\
8 & \textbf{0.92} & 1.17 & 1.82 & 1.39 & 1.30 & 1.18 & 1.48 & 1.33 \\
\hline
\end{tabular*}
\end{table}

\subsubsection{Portfolio-based assessment} \label{sssecportass}
RMSFE-type assessment provides\break a~good diagnostic measure of forecast
performance from a statistical perspective. However, as \citet{bows08}
argued in their paper, in applied economic settings, a pure
error-based assessment measure may fail to fully explain the financial
implications of having used a particular model. Therefore, in this
section we consider an adaptation of the profit based assessment
introduced therein. By using modified versions of their three
trading\vadjust{\goodbreak}
strategies, we create portfolios based on the model forecasts, then
measure the cumulative profit of the strategy. This also serves as a
good capstone exercise for our presentation of the FDFM, as it
simultaneously involves both forecasting \textit{and} curve synthesis:
the primary uses for our model.\looseness=-1

In each strategy we use the same rolling window of 108 months as
described in Section~\ref{sssecmsfeass} so that the trading
algorithm is employed every month over the course of 84 months. Each
period $i$ we create a portfolio consisting of a \$1M purchase of one
bond or set of bonds and a corresponding sale of another bond or set of
bonds for the same amount. Therefore, the net investment per period is
\$0. The decision of which bond to sell and which to buy is made based
on the sign of the predicted spread in their one period returns.

At time $i+1$ we cash out our portfolio and record the cumulative
profit over the 84 month trading period. Denoting the yield at time $i$
of a zero coupon bond of maturity $t$ months as $x_i(t)$, the price of
the bond at time $i$ is
%
\begin{equation} \label{eqbondprice}
P_i(t) = \exp[-tx_i(t)].
\end{equation}
Correspondingly, the price the next period (month) is then
$P_{i+1}(t-1) = \exp[-(t-1)x_{i+1}(t-1)]$ since in the month that has
elapsed the maturity is reduced by, not surprisingly, one month. We
denote the one period return as
%
\begin{equation} \label{eqRt}
R_{i+1}(t) = \biggl[\frac{P_{i+1}(t-1)}{P_i(t)}\biggr] - 1,
\end{equation}
and the log one period return as $r_{i+1}(t) \equiv\ln[1+R_{i+1}(t)]$.
Equations \eqref{eqbondprice} and~\eqref{eqRt} imply
%
\begin{equation} \label{eqrt}
r_{i+1}(t) = tx_i(t) - (t-1)x_{i+1}(t-1).
\end{equation}
Thus, for a forecasted yield $\hat{x}_{i+1|i}(t)$ we have $\hat
{r}_{i+1|i}(t) = tx_i(t) - (t-1)\hat{x}_{i+1|i}(t-1)$, which is a
combination of both actual and forecasted yields. We use the data
presented in Section~\ref{ssecdata} and thus are limited to a set of
nonconsecutive observed maturities. Akin to \citet{bows08}, we rely on
linear interpolation of $x_i(t-1)$ to provide the yield for $x_i(t)$
and use the same random walk forecast (RW) as a benchmark by which to
compare models:
%
\begin{equation} \label{eqRW}
x_{i+1}(t) = x_i (t) + \eta_{i+1}(t),\qquad  \eta_{i+1}(t)
\stackrel
{\mathrm{i.i.d.}}{\sim} \mathit{WN}(0,\nu^2),
\end{equation}
with forecast $\hat{x}_{i+1|i}(t) = x_i (t)$.

\begin{algo}\label{algo1} For this strategy, we adopt the method used in
the second algorithm presented in \citet{bows08}. Ours differs slightly
since the data we use (introduced in Section~\ref{ssecdata}) does not
contain the two month maturity.
Let $T = \{4,5,\ldots,13,16,\ldots,85\}$,
$t_1$ = 4 and $t_{2,j} \in T \setminus\{4\}$; $j=1,\ldots,33$. Every
period $i$ we form a portfolio of sub-portfolios with two bonds $\{
t_1,t_{2,j}\}$. Define weights $w_j$ as the proportion of the
historical absolute excess return on portfolio $\{t_1,t_{2,j}\}$ to the
sum over all $j$ of the same:
\[
w_j = \frac{\sum_{i}|R_{i}(t_{2,j})-R_{i}(t_{1})|}{
\sum_{j}\sum_{i}|R_{i}(t_{2,j})-R_{i}(t_{1})|},
\]
where $i$ spans the period January 1985 to December 1993.

To borrow some notation from \citet{bows08}, let $d_{ij}$ represent the
investment rule for the amount at time $i$ invested in each $j$th
sub-portfolio. To determine the amount invested in each sub-portfolio, let
\[
 d_{ij} = \$1M \times w_j \times\operatorname{sgn}[ \hat
{r}_{i+1|i}(t_{2j}) - \hat{r}_{i+1|i}(t_1)].
\]
We set $d_{ij}= 0$ in the off chance of $\hat{r}_{i+1|i}(t_{2j}) =
\hat
{r}_{i+1|i}(t_1)$. Let $\pi_{i+1}$ denote the time $i+1$ profit
resulting from these rules. Then
\[
\pi_{i+1} = \sum_j d_{ij} [R_{i+1}(t_{2j}) - R_{i+1}(t_1)]
\approx
\sum_j d_{ij} [r_{i+1}(t_{2j}) - r_{i+1}(t_1)].
\]
\end{algo}

The results of this trading strategy are summarized in Table \ref
{tbtradingA2}. Use of the FDFM model results in nearly twice the
cumulative profit produced from the DNS model. Also shown is the
capability of each model in successfully predicting the positive
(1520) and negative (1252) actual spreads of the sub-portfolios in
each period. Surprisingly, the random walk model has the greatest
accuracy in predicting positive spreads (84\%), as compared to the FDFM
(73\%) and DNS (61\%) models. All three models are less accurate in the
prediction of a negative spread, though RW is the worst by far (8\%).
%
\begin{table}
\tabcolsep=0pt
\caption{Algorithm \protect\ref{algo1}: Weighted pairs. Use of the FDFM model results in
nearly twice the cumulative profit produced from the DNS model}\label{tbtradingA2}
\begin{tabular*}{\textwidth}{@{\extracolsep{\fill}}lcccccc@{}}
\hline
& \multicolumn{4}{c}{\textbf{Profit} \textbf{(}$\bolds{\times\$1000}$\textbf{)}} & \multicolumn{2}{c@{}}{\textbf{Directional accuracy of sub-portfolios}}
\\[-6pt]
& \multicolumn{4}{c}{\hrulefill} & \multicolumn{2}{c@{}}{\hrulefill} \\
& & & \multicolumn{2}{c}{\textbf{Percentile}} &$\bolds{+}$&$\bolds{-}$
\\[-6pt]
& & & \multicolumn{2}{c}{\hrulefill} & \multicolumn{2}{c}{} \\
\textbf{Model} & \textbf{Cumulative} & \textbf{Median} &\textbf{10th} & \textbf{90th} &  &  \\
\hline
FDFM & \textbf{1089} & \phantom{00.}5.06 & $-$101.92 & 149.53 & 1102$/$1520 (72.5\%)
& 392$/$1252 (31.3\%) \\
DNS & \phantom{0,}519 & \phantom{00.}5.07 & $-$110.77 & 116.02 & 926$/$1520 (60.9\%) & 538$/$1252
(43\%) \\
RW & \phantom{0}$-$94 & $-$10.52 & $-$190.5\phantom{0} & 163.64 & 1274$/$1520 (83.8\%) & 97$/$1252
(7.7\%) \\
\hline
\end{tabular*}
\end{table}

\begin{algo}\label{algo2}
The strategy in Algorithm~\ref{algo1} is a fairly basic
one: to use \textit{every} available bond at our disposal to predict the
spread between its return and a short-term bond. Our second strategy is
more sophisticated by creating portfolios of an optimal pair of bonds
each period $i$.
Given a fixed value of $t_1$, we choose $t_{2i}$\vadjust{\goodbreak} to optimize the
absolute spread in predicted return each trading period:
%
\begin{equation}\label{eqoptret}
t_{2i} = \arg\max_{t \neq t_1}
|\hat{r}_{i+1|i}(t) - \hat{r}_{i+1|i}(t_1)|.
\end{equation}
\end{algo}

This is an adaption of the third algorithm presented in \citet{bows08}.
There, in a single exercise, the authors fix $t_1 = 3$ and select
$t_{2i}$ according to equation \eqref{eqoptret} each trading
period. Here, we examine multiple choices for $t_1$ and determine
$t_{2i}$ according to equation \eqref{eqoptret} each trading period
for each choice of fixed $t_1$.
Because we examine multiple portfolios, we use a sparser set of
maturities in this exercise than previously, though of the same range.
This set is defined by the observed maturities of Section~\ref{ssecdata}:
\[
t_1,t_{2i} \in T =\{4,7,10,13,16,19,22,25,31,37,49,61,73,85\}.
\]
We perform this exercise for all choices of $t_1$ and $t_{2i}$ as long
as $t_1 < t_{2i}$, and compare the results. Our investment rule $d_i$
at time $i$ and resulting profit~$\pi_{i+1}$ the next period is of a
similar form to Algorithm~\ref{algo1}:
\begin{eqnarray*}
 d_i &=& \$1M \times\operatorname{sgn}[\hat{r}_{i+1|i}(t_{2i}) -
\hat{r}_{i+1|i}(t_1)],\\
\pi_{i+1} &=& d_i [R_{i+1}(t_{2i}) - R_{i+1}(t_1)] \approx
d_i [r_{i+1}(t_{2i}) - r_{i+1}(t_1)].
\end{eqnarray*}
Again, we set $d_{i}= 0$ whenever $\hat{r}_{i+1|i}(t_{2i}) = \hat
{r}_{i+1|i}(t_1)$.
%
\begin{table}[b]
\caption{Algorithm \protect\ref{algo2}: Optimal pairs portfolio}\label{tbtradingA3b}
\begin{tabular*}{\textwidth}{@{\extracolsep{\fill}}lccccccccc@{}}
\hline
& & \multicolumn{3}{c}{\textbf{Profit} \textbf{(}$\bolds{\times\$1000}$\textbf{)}} & \multicolumn{2}{c}{}
& \multicolumn{3}{c@{}}{\textbf{Profit} \textbf{(}$\bolds{\times\$1000}$\textbf{)}}\\[-6pt]
& & \multicolumn{3}{c}{\hrulefill} & \multicolumn{2}{c}{}
& \multicolumn{3}{c@{}}{\hrulefill}\\
& $\bolds{t_1}$ & \textbf{FDFM} & \textbf{DNS} & \textbf{RW} & & $\bolds{t_1}$ & \textbf{FDFM} & \textbf{DNS} & \textbf{RW} \\
\hline
Short & \phantom{0}3 & 1013 & \textbf{3574} & $-$228 & Mid & 21 & \textbf{1246} &
202 & \phantom{0}680 \\
& \phantom{0}6 & 1381 & \textbf{2828} & $-$133 & & 24 & \textbf{1592} & 242 & \phantom{00}70
\\
& \phantom{0}9 & \textbf{1061} & 1013 & $-$297 & & 30 & \textbf{2284} & 203 &
$-$951 \\
& 12 & \textbf{1873} & $-$367 & $-$307 & & 36 & \textbf{1466} & 919 &
$-$173 \\
& 15 & \textbf{1519} & $-$582 & $-$432 & Long & 48 & $-$361 & \textbf{589}
& \phantom{0}236 \\
& 18 & \textbf{1081} & $-$481 & $-$263 & & 60 & \phantom{0}\textbf{740} & 339 & $-$284
\\
& & & & & & 72 & $-$131 & \phantom{0}$-$1 & \phantom{00}\textbf{72} \\
\hline
\end{tabular*}
\end{table}

The results of the strategy are shown in Table~\ref{tbtradingA3b}.
When the choice of $t_1$ is six months or less, the DNS model generates
greater cumulative profit than either of the other models. However,
when the choice of $t_1$ is within 9 and 36 months, the FDFM
consistently generates significantly greater profit than the DNS and RW
models. Thus, when we are free to pick the bond that optimizes the
predicted spread each period, the FDFM performs rather well, provided
the maturity of the first bond is within a certain range. Our final
strategy expands upon this idea.

\begin{algo}\label{algo3}
Because the choice of the optimal second bond can
vary from one period to the next in Algorithm~\ref{algo2},\vadjust{\goodbreak} it is not clear what a
\textit{consistently} good combination is. Thus, for our third
strategy
we consider an exploratory and exhaustive approach as a diagnostic
assessment of with which combination of bonds our model excels. As
such, we expand our set of bonds to include those of longer maturity:
\[
t_1,t_2 \in T =\{
4,7,10,13,16,19,22,25,31,37,49,61,73,85,97,109\}.
\]
\end{algo}

In this modification of strategy 1 from \citet{bows08}, the portfolio
is a simple one consisting of two bonds with maturities $t_1$ and~$t_2$.
For the duration of the strategy, these maturities remain fixed
over all periods $i=1,\ldots, 84$. As before, the decision at time $i$
of which bond to sell and which to buy is made based on the predicted
direction of the spread in log one period returns: $d_i = \$1M \times
\operatorname{ sgn}[ \hat{r}_{i+1|i}(t_2) - \hat{r}_{i+1|i}(t_1)]$ [we set
$d_{i}= 0$ whenever $\hat{r}_{i+1|i}(t_{2}) = \hat{r}_{i+1|i}(t_1)$].
This yields the time $i+1$ profit
\[
\pi_{i+1} = d_i [R_{i+1}(t_2) - R_{i+1}(t_1)] \approx
d_i [r_{i+1}(t_2) - r_{i+1}(t_1)].
\]
We examine the cumulative profit of all combinations of this type or
portfolio such that $t_2 > t_1$.

%
\begin{figure}[b]

\includegraphics{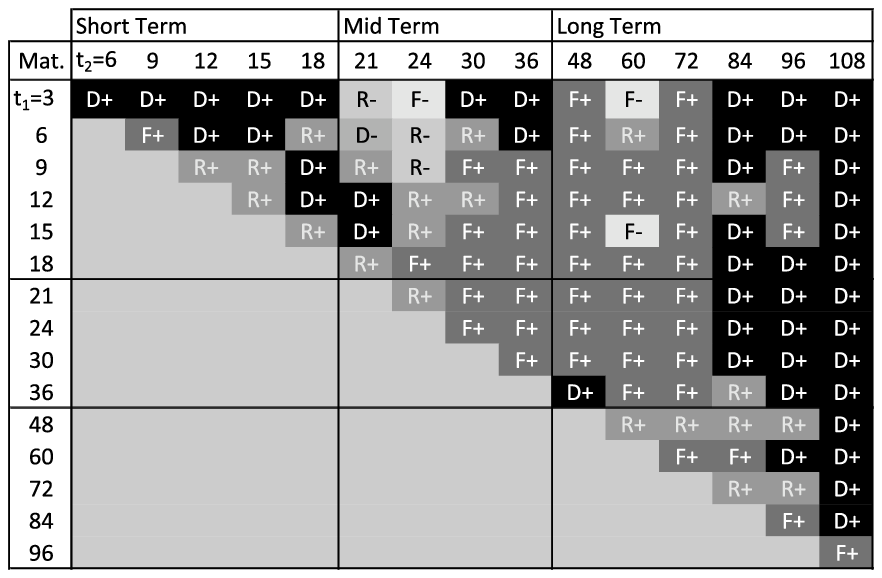}

\caption{Algorithm \protect\ref{algo3}: All combinations of portfolios for $t_2 > t_1$.
The model with the largest cumulative profit is displayed by the first
initial of its acronym with ``$+$'' or ``$-$'' indicating positive or
negative profit.} \label{figtradingA1b}
\end{figure}

Figure~\ref{figtradingA1b} depicts the results of our final trading
strategy. For each combination of $t_2 > t_1$, the name of model with
the largest cumulative profit is displayed in that cell by the first
initial of its acronym (``F'' for FDFM, e.g.). A ``$+$'' or ``$-$'' suffix
indicates the largest profit was positive or negative, respectively.

The FDFM model typically has the greatest profit when $t_2 \in\{
30,\ldots,72\}$. These results are consistent with Sections \ref
{sssecmsfeass} and~\ref{sssecccurveass}: the FDFM was either
comparable or better on RMSFE for forecasting and for imputation on
maturities in this range. We also see a certain similarity in these
results to those of Algorithm~\ref{algo2}. Namely, that the FDFM typically
outperformed the other two models when $t_1$ was exactly in this range.

For the longest maturities ($ > 72$), the DNS model results in greater
profit when $t_1 < 48$. Results for other regions are mixed. Recall
from Section~\ref{ssecdata} that in our data short and mid-term
yields are typically spaced either 3 or~6 months apart, whereas
long-term maturities are spaced 12 months apart. As we saw in Section~\ref{sssecccurveass},
as the spacing between maturities increased,
the FDFM model eventually broke down; it is, after all, very much a
data driven model. DNS, on the other hand, maintains the same factor
loading curves regardless of the data, which could explain its greater
profits at long maturities.

\section{Conclusion and discussion}\label{secconc}
In this paper we developed a method for modeling and forecasting
functional time series. This novel approach synthesizes concepts from
functional data analysis and dynamic factor modeling culminating in a
functional dynamic factor model. By specifying error assumptions and
smoothness conditions for functional coefficients, estimation by the
Expectation Maximization algorithm results in nonparametric factor
loading curves that are natural cubic splines. Thus, for a given time
series of curves we can forecast entire curves as opposed to a discrete
multivariate time series.

The motivating application is yield curve forecasting, where existing
approaches typically exhibit a trade-off of
consistency-with-economic-theory and goodness of fit. However, through
multiple forecasting exercises we show that our model satisfies both
of these criteria. A further online supplement underscores these
results and also showcases the model's viability to settings well
outside of economics and yield curve forecasting and where a prior
theory does not exist. Indeed, this exciting new class of models is
fertile for further development and application.

The present paper focuses on yields of zero coupon bonds. A
particularly interesting direction for future research is the extension
of our modeling framework to yields implied by commodities that include
convenience factors. For example, see \citet{casassus2005stochastic}
and \citet{chua2008dynamic}. A potential difficulty in this regard is
the consistent control of multiple parameters: the commodity, the
maturity and the liquidity of said maturity, for example. Another
interesting direction of research is to develop nonlinear time series
models for functional data; existing approaches for nonlinear
univariate time series modeling [see \citet{fan2005nonlinear},
Section~1.5.4] may be helpful for that purpose.

\section*{Acknowledgments}
The authors would like to express their sincere gratitude to the
Editor, Associate Editor and reviewers whose comments have helped to
refine and improve the scope and presentation of the paper.

\begin{supplement}\label{suppA} 
\stitle{Simulation studies and technical proofs}
\slink[doi]{10.1214/12-AOAS551SUPP} 
\slink[url]{http://lib.stat.cmu.edu/aoas/551/supplement.pdf}
\sdatatype{.pdf}
\sdescription{The online supplement contains the following: (1)
additional simulation studies to further illustrate the advantages of
our method; (2) detailed proofs of Theorem~\ref{thcv,gcvthm} and Propositions
\ref{thSx}--\ref{prop4}.}
\end{supplement}



\printaddresses

\end{document}